\documentclass[preprint]{aastex}

\newcommand{\ie}{{\em i.e.},\ }
\newcommand{\eg}{{\em e.g.},\ }
\newcommand{\setal}{{\em et al.}\ }
\newcommand{\etal}{{\em et al.}}

\newcommand{\civ}{C\,{\sc iv}}
\newcommand{\ebv}{$E(B-V)$}

\newcommand{\ha}{H$\alpha$}

\newcommand{\kms}{km\,s$^{-1}$}

\newcommand{\mgii}{Mg\,{\sc ii}}

\newcommand{\alii}{Al\,{\sc ii}}

\newcommand{\aliii}{Al\,{\sc iii}}
\newcommand{\AlIII}{Al\,{\sc iii}\,$\lambda\lambda$1854,1862}

\newcommand{\ciii}{C\,{\sc iii}]}
\newcommand{\CIII}{C\,{\sc iii}]\,$\lambda$1908}

\newcommand{\crii}{Cr\,{\sc ii}}
\newcommand{\fei}{Fe\,{\sc i}}
\newcommand{\feii}{Fe\,{\sc ii}}
\newcommand{\feiii}{Fe\,{\sc iii}}
\newcommand{\hei}{He\,{\sc i}}
\newcommand{\HeI}{He\,{\sc i}\,$\lambda$3188}

\newcommand{\HeIIsf}{He\,{\sc ii}\,$\lambda$1640}

\newcommand{\MgI}{Mg\,{\sc i}\,$\lambda$2852}

\newcommand{\mnii}{Mn\,{\sc ii}}

\newcommand{\Nv}{N\,{\sc v}}

\newcommand{\NIii}{Ni\,{\sc ii}}

\newcommand{\Si}{S\,{\sc i}}

\newcommand{\SIii}{Si\,{\sc ii}}

\newcommand{\SIiv}{Si\,{\sc iv}}

\newcommand{\znii}{Zn\,{\sc ii}}

\slugcomment{Accepted by the Astronomical Journal}

\shorttitle{Peculiar BAL Quasars}
\shortauthors{Brunner, R. J. \etal}

\begin{document}


\title{Peculiar Broad Absorption Line Quasars found in DPOSS\footnote{Some of the data presented herein were obtained at the W.M. Keck Observatory, which is operated as a scientific partnership among the California Institute of Technology, the University of California and the National Aeronautics and Space Administration. The Observatory was made possible by the generous financial support of the W.M. Keck Foundation.}}


\author{Robert J. Brunner\altaffilmark{1}}
\affil{Palomar Observatory, California Institute of Technology, Pasadena, CA 91125}
\email{rb@astro.caltech.edu}

\author{Patrick B. Hall}
\affil{Pontificia Universidad Cat\'{o}lica de Chile, 
Departamento de Astronom\'{\i}a y Astrof\'{\i}sica, 
Facultad de F\'{\i}sica, Casilla 306, Santiago 22, Chile,
and Princeton University Observatory, Princeton, NJ 08544-1001}

%\altaffiltext{}{}

\author{S. George Djorgovski, R.R. Gal\altaffilmark{2}, A.A. Mahabal, P.A.A. Lopes,  
R.R. de Carvalho\altaffilmark{3}, S.C. Odewahn\altaffilmark{4}, S. Castro\altaffilmark{5}, D. Thompson}
\affil{Palomar Observatory, California Institute of Technology, Pasadena, CA 91125}

\author{F. Chaffee}
\affil{W.M. Keck Observatory, 65-1120 Mamalahoa Highway, Kamuela, HI 96743
}

\author{J. Darling\altaffilmark{6}, and V. Desai\altaffilmark{7}}
\affil{Palomar Observatory, California Institute of Technology, Pasadena, CA 91125}

\altaffiltext{1}{Current address: Department of Astronomy, University of Illinois, Urbana, IL, 61801}

\altaffiltext{2}{Current address: Department of Physics and Astronomy, 
Johns Hopkins University, Baltimore, MD 21218}
\altaffiltext{3}{Current address: Observatorio Nacional, CNPq, Rio de Janeiro, Brasil}
\altaffiltext{4}{Current address: Department of Physics and Astronomy, Arizona State University, 
Tempe, AZ 85287}
\altaffiltext{5}{Current address: Infrared Processing and Analysis Center, 770 S. Wilson Ave, Pasadena, CA, 91125}
\altaffiltext{6}{Current address: Department of Astronomy, Cornell University, Ithaca, NY 14853}
\altaffiltext{7}{Current address: Department of Astronomy, University of Washington, Seattle, WA 98195}


\begin{abstract}

With the recent release of large (i.e., $ \ga$ hundred million objects), well-calibrated photometric 
surveys, such as DPOSS, 2MASS, and SDSS, spectroscopic identification of  important targets is no longer a simple issue. In order to enhance the returns from a spectroscopic survey, candidate sources are often preferentially selected to be of interest, such as brown dwarfs or high redshift quasars. This approach, while useful for targeted projects, risks missing new or unusual species. We have, as a result, taken the alternative path of spectroscopically identifying interesting sources with the sole criterion being that they are in low density areas of the $g - r$ and $r - i$ color-space defined by the DPOSS survey. In this paper, we present three %new 
peculiar broad absorption line quasars that were
discovered during this spectroscopic survey, demonstrating the efficacy of this approach.
PSS J0052+2405 is an  Iron LoBAL quasar at a redshift $z = 2.4512 \pm 0.0001 $ with very broad absorption from many species. PSS J0141+3334 is a reddened LoBAL quasar at $z = 3.005 \pm 0.005$ with no obvious emission lines. PSS J1537+1227 is a Iron LoBAL at a redshift of $z = 1.212 \pm 0.007$ with strong narrow \mgii\ and \feii\ emission. Follow-up high resolution spectroscopy of these three quasars promises to improve our understanding of BAL quasars. The sensitivity of particular parameter spaces, in this case a two-color space, to the redshift of these three sources is dramatic, raising questions about traditional techniques of defining quasar populations for statistical analysis.

\end{abstract}




\section{Introduction}

%Talk about quasars. Why are they interesting. What do we need to learn.
%Why are BALs and more so Extreme BALs important.
Quasars have been studied for forty years now (\eg Sandage 1965), but our understanding of them
is still sketchy in many ways.  The currently accepted paradigm is that quasars derive
their luminosity from accretion onto supermassive black holes (see, \eg Small \& Blandford 1992).
However, the detailed structure of the central engine is a mystery:
we do not know exactly how matter accretes onto the black hole, though an
accretion disk seems likely; we do not know why some quasars have strong jets
(or how they are formed); and we do not know exactly how broad absorption line
(BAL) outflows fit into the overall quasar model, even though the mass loss
rates in such outflows could be comparable to the overall accretion rates (\eg Scoville \& Norman 1995).

%BAL quasars have been studied for almost 35 years \citep{lyn67}
BAL quasars show absorption from gas with blueshifted
outflow velocities of typically 0.1$c$ (for a good overview of BAL quasars see Weymann 1995).
Most known BAL quasars are HiBALs, with absorption only from high-ionization species
like \civ, but about 15\% are LoBALs, which also show absorption from
low-ionization species like \mgii.  The rare Iron LoBALs, or FeLoBALs,
also show absorption from excited fine-structure levels or excited atomic terms
of \feii\ or \feiii.

Recent spectroscopic surveys for quasars, such as the two degree field survey (2DFQRS; Boyle \setal 1999) and the Sloan Digital Sky Survey (SDSS; Schneider \setal 2002), have dramatically increased the number of spectroscopically confirmed quasars. As a direct result, the number of known BALs, and in  particular, LoBALs and FeLoBALs, has increased as well, enabling the identification of peculiar quasars, whose BAL outflows show properties
never before seen.  Due to their peculiar nature, detailed studies of these quasars often provide important insight into the physical characteristics of BAL outflows, and quasar models in general. Before high-resolution spectroscopy can be employed on these systems, however, these peculiar quasars must be found.

Due to their rare nature, identifying candidates of such objects requires large photometric surveys, such as the FIRST radio survey (Becker \setal 1997), the SDSS (York \setal 2000),
and the Digitized Palomar Observatory Sky Survey (DPOSS; Djorgovski \setal 2001). From the catalogs that are generated from these surveys, different selection criteria have been employed to pick suitable candidates for spectroscopic follow-up. 

In this paper we present three peculiar BAL quasars that were discovered during a spectroscopic follow-up of color-space outliers from the DPOSS survey.
We outline the observations used in \S\,2, analyze the spectra in \S\,3,
discuss some implications of these objects in \S\,4, and summarize our
conclusions in \S\,5.

\section{Observations}

The quasars presented in this paper were initially targeted due to their location in the $g - r$ and $r - i$ color space as quantified by the Digitized Palomar Observatory Sky Survey. The DPOSS survey is detailed extensively elsewhere (Djorgovski \setal 2001, and references therein). In order to clarify the selection of these three sources, however, the rest of this section provides a brief overview of DPOSS.

DPOSS is based on the POSS-II photographic survey (Reid \setal 1991), which covers the Northern sky ($\delta > -3\degr$). The POSS-II plates were obtained at the 48-inch Oschin Scmidt telescope at Palomar in three bands: blue-green, IIIa-J + GC395, red, IIIa-F + RG610, and near-IR, IV-N + RG9. The plates were obtained following a strategy based on 897 fields, where each field is approximately $6.5\degr$ on a side (\ie the size of an individual plate) while the individual field centers are spaced $5\degr$ apart. As a result, nearly half of the total survey area is covered by more than one plate, improving the overall calibration.

The photographic plates were digitized using a modified PDS scanner at STScI (Lasker \setal 1996) producing a digital image that is 23,040 pixels square, with 1$\arcsec$ pixels. Each image file is approximately one Gigabyte in size, and the total survey is nearly three Terabytes. The digital image files for all scans $\delta > 15\degr$ were processed using the SKICAT software (Weir \setal 1995), producing a catalog of approximately 60 parameters for each object. Object classification was performed using a subset of these parameters as described in Odewahn \setal (2003).

\begin{figure}
\epsscale{0.75}
\plotone{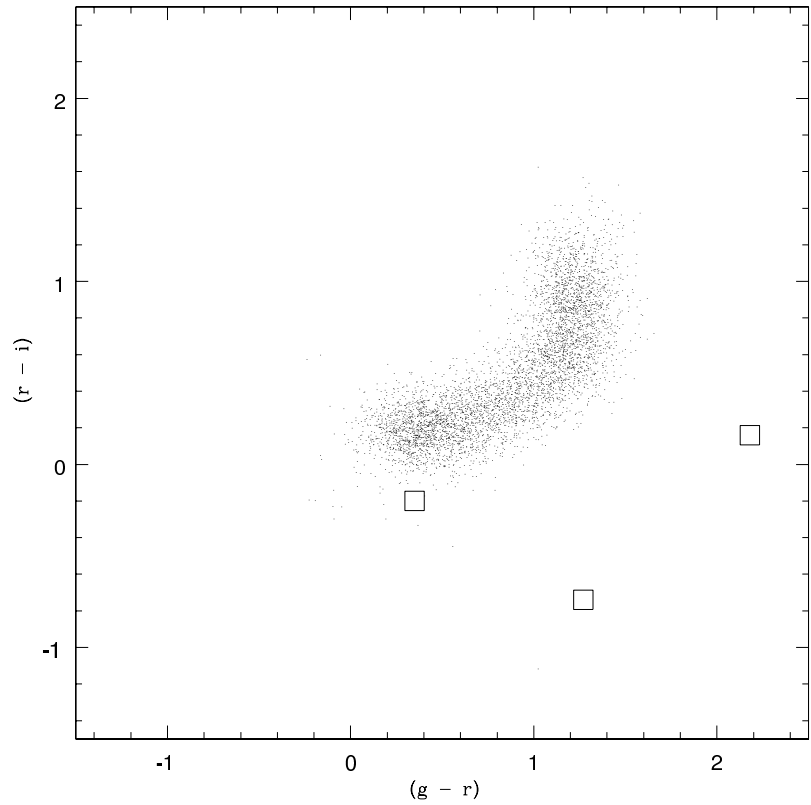}
\caption[]{The color space used to select color-outliers. The three quasars presented in this paper are marked by the large squares. The small dots are several thousand stellar sources, with similar magnitudes to the quasars presented herein, extracted randomly from the survey.}\label{target}
\end{figure}

The photographic catalog data was calibrated using CCD calibration data in  the $g$, $r$, and $i$ filters (Gal \setal 2003). A second correction was applied to the catalog to correct for plate vignetting (Mahabal \setal 2003). Finally, extinction corrections were applied using the Schlegel, Finkbeiner, \& Davis (1998) prescription. The typical limiting magnitudes for the calibrated catalog data are $g_J \sim 20.5^m$, $r_F \sim 20.7^m$, and $i_N \sim 20.3^m$. The three sources presented in this paper are shown in Figure~\ref{target}, along with random stellar sources with similar magnitudes to the quasars presented herein. In addition, the coordinates, magnitudes and measured redshifts for these sources are provided in Table~\ref{dataTable}.

\subsection{Optical Spectroscopy}

Discovery spectra of all three objects were obtained at the Palomar Observatory
200-inch Hale telescope, using the Double Spectrograph (DBSP) instrument
(Oke \& Gunn 1982). All observations used a 2$\arcsec$ wide, long slit, and in all cases the slit PA
was close to the parallactic.  Exposures of standard stars from
Oke \& Gunn (1983) were used to remove the instrument response function and provide at least a
rough flux calibration, and exposures of arc lamps were used to derive the
wavelength solutions.  All data were processed using standard techniques.

Observations of PSS J0052+2405 were made on 02 and 04 September 1997 UT, in
non-photometric conditions.  A set of exposures with integration times of 200, 
400, and 1200 s were obtained.  We used  
a 300 l/mm grating on the blue side of the DBSP, giving 
a dispersion of 2.17 \AA/pixel and a FWHM resolution of 11 \AA, covering 
the wavelength range from $\sim$ 3360 \AA\ to 6860 \AA.  On the red side we
used a 316 l/mm grating giving a
dispersion of 3.06 \AA/pixel and a resolution of 11 \AA, covering
the wavelength range $\sim$ 6760 \AA\ to 9290 \AA.  A dichroic with a split
wavelength near 6800 \AA\ was used.

Observations of PSS J0141+3334 were obtained on 10 November 1999 UT, in good
conditions; exposure times were 120 and 800 s.
We used gratings with 600 l/mm (blue) and 158 l/mm (red), covering the
wavelength ranges $\sim$ 3400 to 5200 \AA\ (blue), and $\sim$ 5120 to 10100
\AA\ (red), and a 5200 \AA\ dichroic. This is the discovery spectrum shown in Figure~\ref{01413334Spec} and used in the analysis presented in this paper.

Observations of PSS J1537+1227 were obtained on 21 May 1996 UT, in mediocre
conditions, with exposures of 300 and 900 s.
The same gratings, but with a 5500 \AA\ dichroic, and the wavelength coverage
$\sim$ 3900 to 5600 \AA\ (blue side) and $\sim$ 5550 to 8070 \AA\ (red side).
Additional data were obtained on 12 May 1999 UT, but were affected by
instrument problems and not used.

Follow-up observations of PSS J0052+2405 were obtained at the WMKO Keck-I 10-m
telescope on 04 October 1997 UT, using the Low Resolution Imaging Spectrometer
(LRIS; Oke et al. 1995), 
in good conditions.  A single 1800 s integration was
obtained using a 600 l/mm grating centered at $\lambda \sim 6000$ \AA, and
covering a wavelength range of $\sim 4800$ to $\sim 7300$ \AA, with a 
dispersion of 1.25 \AA/pixel and a FWHM resolution of $\sim 9$ \AA\ through
a 1.5$\arcsec$ wide slit.  These data were superseded by a moderate-resolution
spectrum obtained at the Keck-II 10-m telescope on 30 December 1999 UT,
using the Echellette Spectrograph and Imager
(ESI; Sheinis et al. 2002).
Exposures of 900 and 1500 s were obtained, in good conditions.  The instrument
resolution is a constant 11.4 km/s/pixel or FWHM $\sim 74$ km/s through the
1.0 arcsec slit, and it covers nearly the entire visible light window, from
$\sim 3900$ \AA\ to $\sim 11000$ \AA.
This is the discovery spectrum shown in Figure~\ref{00522405Spec} and used in the analysis presented in this paper.

Follow-up observations of PSS J1537+1227 were obtained using LRIS as follows.
On 02 April 1998 UT, we obtained two exposures of 600 s, using a 400 l/mm
grating covering a wavelength range from $\sim 5850$ \AA\ to $\sim 9550$ \AA,
giving a resolution of FWHM $\sim 13$ \AA\ through a 1.5 arcsec wide slit.
On 03 April 1998 UT, we obtained two exposures of 600 s, using a 300 l/mm
grating covering a wavelength range from $\sim 3900$ \AA\ to $\sim 9000$ \AA,
giving a resolution of FWHM $\sim 17$ \AA\ through a 1.5$\arcsec$ wide slit.
On 12 June 1999 UT, we obtained an additional two exposures of 600 s and two
of 900 s using the same grating and wavelength coverage, but with a 1 arcsec
wide slit, giving a resolution of FWHM $\sim 12$ \AA.  In addition, we also
obtained two exposures of 1200 s using a 600 l/mm
grating covering a wavelength range from $\sim 5900$ \AA\ to $\sim 8450$ \AA,
giving a resolution of FWHM $\sim 6$ \AA\ through a 1 arcsec wide slit.
The April 1998 data were obtained in photometric conditions, but the June 1999
data were taken through a thin cirrus.  All of the LRIS data were reduced and
calibrated in a standard manner. This is the discovery spectrum shown in Figure~\ref{15371227Spec} and used in the analysis presented in this paper.

\subsection{Near-Infrared Spectroscopy}

An $H$-band spectrum of PSS J1537$+$1227 was obtained on UT 1999 August 21 
using the facility near-infrared spectrograph NIRSPEC (McLean \setal 1998)
on the Keck\,{\sc II} telescope.  The NIRSPEC detector 
is an ALADDIN 1024$^2$ pixel InSb array.  The low-resolution mode was used 
with an 0\farcs57 wide (4 pixel) slit, giving a spectral resolution of $R 
\sim 1500$.  Two exposures of 600\,s were obtained on the source, dithered 
14$\arcsec$ along the slit.  Two exposures of SAO\,101725, an 
A2{\sc IV} star, were also obtained at similar airmass to remove the effects 
of atmospheric absorption.  Domeflats were used to correct for the 
pixel-to-pixel variations in detector quantum efficiency.

Low-resolution spectra from NIRSPEC are rotated by $\sim 5\degr$, as well 
as having significant distortions relative to perfect alignment with the 
rows and columns of the detector.  To reduce the data, we first identified 
and interpolated over the obvious bad pixels and cosmic rays in each frame 
so they would not be smeared out by further processing.  The data were then 
flatfielded with a normalized domeflat and the spectra rotated and chopped 
to cover the area spanned by the 42$\arcsec$ long slit.  Each frame 
was corrected using a distortion map made from the night sky OH emission 
lines.

Individual spectra were extracted across an 11 pixel (1\farcs6) window
using variance weighting in the IRAF\footnote{IRAF is distributed by the National Optical Astronomy Observatories, which are operated by the Association of Universities for Research in Astronomy, Inc., under cooperative agreement with the National Science Foundation.} {\sc apextract} package.
Wavelength calibration was determined from the night sky OH emission
lines, which show a residual rms scatter of 0.5\,\AA.  The atmospheric
extinction was removed by dividing the spectra by the standard star
normalized by a 9730\,K blackbody spectrum, corresponding to the
standard's A2{\sc IV} spectral type, covering the same wavelength
range.  The two spectra of PSS J1537$+$1227 were then averaged. The final near infrared spectrum of PSS J1537$+$1227 is shown in Figure~\ref{pss1537_Hspec}.

\section{Peculiar BAL Quasars}

In this section we provide an individual analysis for each of the
three peculiar BAL quasars presented in this paper.  Identification of the
absorption lines (using Moore 1950 and Moore 1962)
and analysis of the individual spectra follow the
techniques outlined in Hall \setal 2002.  In particular, for all objects we
quote both the traditional balnicity index (BI; Weymann \setal 1991) and 
the restrictive absorption index (AI; Hall et al. 2002).  The AI is designed to
include objects with troughs which are relatively narrow or close to the quasar
redshift, since detailed studies show that such troughs often %otherwise 
share many of the characteristics of BAL troughs.

\subsection{PSS J0052+2405}

PSS J0052+2405 is an FeLoBAL quasar with very broad absorption from many species, presented in Figure \ref{fig0052opt}.
We adopt $z = 2.4512 \pm 0.0001$ from narrow \SIiv\ and \AlIII\ absorption.
The long-wavelength end of our spectrum is a trough from \feii+\mgii;
moving toward shorter wavelengths,
the spectrum recovers before encountering another \feii\ trough near 2400\,\AA,
recovers again near \CIII\ before encountering \AlIII\ absorption,
and finally shows a weak recovery at \SIiv.
\civ\ is not prominent because of overlying absorption from \feii\ multiplets
UV42-46, similar to SDSS~0437$-$0045 (Hall \etal\ 2002).  

Many species are detected at the peak absorption redshift of $z = 2.406$,
including \SIiv, \SIii, \civ, \alii, \aliii, \NIii, \znii, \crii, \mnii,
and numerous \feii\ multiplets (including UV42-43, which are likely responsible
for the trough at the expected wavelength of \HeIIsf).  %along with UV68
The highest numbered \feii\ multiplets readily identifiable are UV79 and UV80,
which arise from terms $\sim$1.7\,eV above ground.  
%However, the trough at
%$\sim$2220\,\AA\ is unidentified and may include a contribution from UV168
%($\sim$2.7\,eV above ground), as discussed in \S5.1.3 of Hall \etal\ (2002).
%	and possibly \fei\ and \SIi.
%	The latter two are uncertain identifications,
%	as neutral absorption is rare in BAL quasars and there is no sign of 
%	other neutral absorption (e.g. \mgi\,$\lambda$1827).
%	%	There may be C\,{\sc i} at $z = 2.4512$, but it's very subtle.
%%Also, $z=2.42$ is required for \SIi\,$\lambda$2207 to match the 7550\,\AA\ 
%%trough, and there is no evidence for other absorption at that redshift.
%And at $z=2.406$, \fei\,$\lambda$2167 is a good fit to the 7380\,\AA\ trough
%but \SIi\,$\lambda$2207 is not a good fit to the 7550\,\AA\ trough.
%Considering excited level absorption from the same multiplets as those lines
%improves the fit for \SIi\ but worsens it for \fei.  
%Thus, whatever the conditions in the putative neutral BAL component,
%the wavelengths of such absorption are not a good fit to the observations.
%
\znii\ and \crii\ are comparably strong at $z = 2.406$ in this object,
indicating little dust.
What appears to be additional \znii\ at slightly lower $z=2.39$ without
accompanying \crii\ is in fact residuals from telluric absorption.

%Identifications are uncertain for the troughs at observed wavelengths
%%UV38,84?	1700	5800\,\AA\ (narrow), 
%UV99,100?	1775	6050\,\AA, 
%UV83,94-96?,122-123?,169-170?,186-187?,228?	1950-2000	6800\,\AA, 
%UV168?		2220		7550\,\AA.
%(all observed wavelengths).
%\feii\ multiplets could possibly explain all these troughs, though it would
%require absorption from levels at least 2.7\,eV above ground.

There are only two or three possible continuum windows in our spectrum,
so it is impossible to say how much this object might be reddened.  We do
roughly estimate the BI and AI values from the \alii\ trough using the 
continuum shown in Figure \ref{fig0052opt}.  
We measure BI=1090\,\kms\ and AI=4240\,\kms.  The absorption extends to a
velocity of at
least 7180\,\kms, beyond which there is confusion with a \SIii\ trough.

\subsection{PSS J0141+3334}

PSS J0141+3334 is a reddened LoBAL quasar at $z = 3.005 \pm 0.005$, as
estimated from the red edges of the \SIiv, \civ\ and \alii\ absorption troughs, seen in Figure~\ref{fig0141opt}.
This may be a slight underestimate, as the \aliii\ and possibly the \Nv\ 
troughs may set in at slightly higher redshift.
\alii\ has its peak absorption depth at $z = 2.962$,
while the peak depth of \aliii\ is at higher outflow velocity.
There appears to be another trough at 1430\,\AA\ (assuming $z = 3.005$), 
but its identification is unclear.  A \Si\ identification would be surprising
given the lack of absorption from other neutral species which are not
shielded from ionization by \ion{H}{1}.  There is also narrow intervening 
\feii\ absorption in the spectrum at $z = 2.422$ and \mgii\ absorption
at $z=2.3325$; the latter system may also show weak \feii\ absorption.

The steep continuum of PSS J0141+3334 below $\sim$2000\,\AA\ rest frame is
likely due to dust reddening,  because there are no plausible transitions which
could blanket the spectrum with absorption shortward but not longward of \aliii.
The spectrum just longward of \civ\ at 1550-1600\,\AA\
is probably close to the true continuum level,
rather than being the bottom of an extended \alii\ trough.

We adopt the continuum shown in Figure \ref{fig0141opt}. There may be weak broad \civ\ emission above this continuum, and broad \ciii\ as well, but in general the broad emission in this object appears weak or absorbed. We estimate a SMC extinction law color excess \ebv=0.30$\pm$0.03 from comparison of our adopted  continuum with the composite SDSS quasar of Vanden Berk \setal (2001). We also measure BI=10230\,\kms\ and AI=13800\,\kms\ from the \civ\ trough using this continuum and a limiting velocity of 20410\,\kms\ to avoid confusion with the 1430\,\AA\ trough.  

\subsection{PSS J1537+1227}

PSS J1537+1227 
is a FeLoBAL with strong narrow \mgii\ and \feii\ emission, presented in Figure \ref{fig1537opt};
a less extreme example is FIRST~0840+3633 (Becker \setal 1997).
It has a redshift of $z = 1.212 \pm 0.007$
as determined from the broad \ha\ line in its $H$-band spectrum, shown in Figure \ref{fig1537ir}.
\ha\ is no doubt blended with [\ion{N}{2}], but there is no sign of
[\ion{S}{2}] or other, weaker lines in the spectral range covered.
The rise at the short wavelength end of the $H$-band spectrum is
probably not real, as it does not match any known strong line or \feii\ blend.

%PSS J1537+1227 has Mg{\sc ii} and Fe{\sc ii}~UV1 emission (dot-dashed lines)
%which is much narrower than the \ha\ emission.
%The redshift from Mg{\sc ii} is $z = 1.2025 \pm 0.0005$, which is a blueshift
%of 1290$\pm$950\,\kms\ from the \ha\ redshift.
%PSS J1537+1227 shows extensive ultraviolet absorption from \feii\ 
%The absorption troughs %covers $1.17 \leq z \leq 1.199$ (3980\,\kms) and 
%appear deepest at $z=1.187$.
%Blueward of \mgii, the spectrum includes absorption from
%\SIii, \aliii, \feiii\,UV34 (near 1910\,\AA), 
%but not \feiii\,UV48 (near 2070\,\AA), which is typical,
%\crii+\znii, and numerous \feii\ multiplets (dotted lines)
%up to at least UV79 and UV80 ($\sim$~1.7\,eV above ground; \cf PSS J0052+2405).
%The unidentified $\sim$2220\,\AA\ trough present in PSS J0052+2405 is present
%here as well, indicating the possible presence of \feii\ absorption from terms
%$\sim$~2.7\,eV above ground.
%
%Absorption from \HeI\ and \hei\,$\lambda$2946 may also be present,
%and possibly weak \MgI.
%perhaps even metastable excited \mgii\,$\lambda$2934 (Grandi \& Phillips 1978).
%These features are useful diagnostics of physical conditions in the
%BAL gas, but a higher resolution spectrum is needed to sort out 
%which of them are in fact present.

PSS J1537+1227 has Mg{\sc ii} and Fe{\sc ii}~UV1 emission (dot-dashed lines)
which is much narrower than the \ha\ emission.
The redshift from Mg{\sc ii} is $z = 1.2025 \pm 0.0005$, which is a blueshift
of 1290$\pm$950\,\kms\ from the \ha\ redshift.
The absorption troughs %covers $1.17 \leq z \leq 1.199$ (3980\,\kms) and 
appear deepest at $z=1.187$, a further blueshift of 2120\,\kms.
Blueward of \mgii, the spectrum includes absorption from
\SIii, \aliii, \feiii\,UV34 (near 1910\,\AA), 
%but not \feiii\,UV48 (near 2070\,\AA), which is typical,
\crii+\znii, \NIii\ and numerous \feii\ multiplets (dotted lines)
up to at least UV79 and UV80 ($\sim$1.7\,eV above ground; cf. PSS J0052+2405).
%
%The unidentified $\sim$2220\,\AA\ trough present in PSS J0052+2405 is present
%here as well, indicating the possible presence of \feii\ absorption from terms
%$\sim$2.7\,eV above ground.
%
Absorption from \HeI\ and \hei\,$\lambda$2946 may also be present,
and possibly weak \MgI.
%perhaps even metastable excited \mgii\,$\lambda$2934 (Grandi \& Phillips 1978).
These features are useful diagnostics of physical conditions in the
BAL gas, but a higher resolution spectrum is needed to sort out 
which of them are in fact present.

It is difficult to know where to place the continuum, since the spectrum is a
complex blend of \feii\ emission and absorption even longward of \mgii. 
There is certainly some \feii\,UV60,61,78 absorption at 2860-3000\,\AA\ 
rest frame, since absorption from multiplets up to UV80 and Opt7
are identifiable elsewhere in the spectrum.  %present shortward of \mgii.  
In Figure \ref{fig1537opt} we use dotted lines to show the wavelengths of these
strong absorption lines of excited \feii, and dot-dashed lines to show emission
from multiplets UV1,60,61,78 and Opt1,6,7.  
%This emission is plotted at the
%longest-wavelength lines of the multiplets, except for the semi-forbidden
%multiplets UV60,61 and Opt1,
%for which the wavelengths of the strongest lines are used.

%The good agreement of the dot-dashed lines with the emission peaks and the
%dashed lines with the absorption shows that the spectrum can be understood
%as emission at $z=1.2025$ coupled with absorption, which peaks at $z=1.187$
%($\Delta v=2120$\,\kms).  
%
%We note that the relative strengths of the UV1, UV61, UV60+78, and Opt7 peaks
%may be in better agreement with theoretical models than the spectrum of
%QSO~2226$-$3905 studied in de Kool \setal (2002).  Contrary to their suggestion
%of a problem with the theory, it may just be that the \feii\ emission line
%regions span a wider range of physical conditions than considered in their
%modeling.  This is merely a suggestion, however, since the relative strengths
%of the \feii\ lines in this object are affected by absorption.  In particular,
%the UV2 and UV62,63 peaks are much weaker here than in theoretical models.
%More detailed modelling based on higher resolution spectra is needed to
%determine whether this is a problem with the theory or whether it can be
%explained by overlying absorption.

The good agreement of the dot-dashed lines with the emission peaks and the
dashed lines with the absorption shows that the spectrum can be understood as
emission at $z=1.2025$ coupled with absorption which is strongest at $z=1.187$.
We note that the relative strengths of the UV1, UV61, UV60+78, and Opt7 \feii\
emission peaks may be in better agreement with theoretical models than the spectrum
of QSO~2226$-$3905 studied in de Kool \etal\ 2002.  Contrary to their suggestion
of a problem with the models, it may just be that the \feii\ emission line
regions span a wider range of physical conditions than considered in their
modeling.  However, it is also true that the relative strengths of the
\feii\ lines in PSS J1537+1227 are affected by absorption.  In particular,
the UV2 and UV62,63 peaks are much weaker here than in theoretical models.
More detailed modelling based on higher resolution spectra is needed to
determine whether this is a problem with the theory or whether it can be
explained by overlying absorption.

%There is in fact at least a weak absorption trough in this object that
%hints that the absorption may be even more complicated than it appears.  
%The $\sim$2480\,\AA\ trough might be explained as \feii\ absorption if terms
%$\gtrsim$~2.7\,eV above ground are populated, as suggested by the 2220\,\AA\ 
%trough (see above).  However, even if this is the case, absorption from the
%relevant multiplets (\feii\,UV148,161,179) is a poor fit to the 2480\,\AA\ 
%trough.
%A much better fit is provided by \fei\,UV9.  However, there is no corresponding
%trough of \fei\,UV1 at 2970\,\AA, which should be at least as strong.  In fact,
%there is a local maximum at the expected wavelength of \fei\,UV1.  \feii\ 
%emission could conceivably hide a \fei\,UV1 trough, but that would require
%that the true peaks of the emission lines be considerably higher than observed.
%
%Nonetheless, we cannot rule out the presence of \fei\ without a higher
%resolution spectrum.  Such spectroscopy would constrain the physical parameters
%of the BAL outflow even if no \fei\ is present (\eg de Kool \setal 2002), but 
%would do so even more tightly if it is.  This is because \fei\ has an
%ionization potential less than that of \ion{H}{1}, and so is only found in
%gas with a low ionization parameter, which makes it a very useful diagnostic.

There is in fact at least a weak absorption trough in this object that
hints that the absorption may be even more complicated than it appears.  
The $\sim$2480\,\AA\ trough might be explained as \fei\,UV9.  
However, there is no corresponding
trough of \fei\,UV1 at 2970\,\AA, which should be at least as strong.  In fact,
there is a local maximum at the expected wavelength of \fei\,UV1.  \feii\ 
emission could conceivably hide a \fei\,UV1 trough, but that would require
that the true peaks of the emission lines be considerably higher than observed.
Nonetheless, we cannot rule out the presence of \fei\ without a higher
resolution spectrum.  Such spectroscopy would constrain the physical parameters
of the BAL outflow even if no \fei\ is present (e.g., de Kool \etal\ 2002), but 
would do so even more tightly if it is.  This is because \fei\ has an
ionization potential less than that of \ion{H}{1}, and so is only found in
gas with a low ionization parameter, which makes it a very useful diagnostic.

For now, since \fei\ is rarely seen in BAL outflows, we conservatively
assume that only \feii\ is present in this object, which allows us to make
an educated guess about the continuum level.
There are no strong \feii\ transitions immediately redward of 2800\,\AA\ and
3000\,\AA, so we adopt those as continuum windows, along with regions near
3350\,\AA\ and 3600\,\AA\ which are also free of strong \feii\ transitions
from terms $\lesssim$~4\,eV above ground.  
Adopting the continuum shown in Figure~\ref{fig1537opt}, we find from the
\mgii\ trough that AI=1780\,\kms\ but that the BI is only marginally nonzero
(3\,\kms) since the \mgii\ absorption trough only extends 
$\sim$5035\,\kms\ shortward of the systemic redshift.

\section{Discussion}

The three peculiar BAL quasars presented in this paper were selected from the DPOSS survey (Djorgovski \setal 2001) due to their unusual locations within the $g-r$ versus $r-i$ color space. While large, multi-band photometric surveys, such as DPOSS, provide great opportunities for improving the statistical quantification of different astrophysical source classes, the challenge of finding novel sources is equivalent to finding needles in a haystack. When the number of detected sources in surveys of this type exceed one hundred million, even well-intentioned color selections can fail to uncover  possible interesting sources, and the resulting selection effects can be difficult---if not impossible---to completely characterize.

To demonstrate this difficulty, we examined the locus traced by these quasars in the target color space as their observed emission redshift was synthetically shifted ($\Delta z = \pm 0.1$). This process involved shifting each individual spectrum, both blueward and redward, and convolving the resulting spectrum with the $g$, $r$, and $i$ digital filter curves, obtained from the Palomar Observatory website. In several cases, the shifted spectrum did not completely cover the same wavelength range as one of the two end filters (\ie $g$ or $i$). To account for this, the spectrum was extended across the necessary wavelength range, by extrapolating the last flux value. In no case did this small approximation make any significant contribution to the resulting flux measurements (which is obvious since the filters fall sharply at their edges).

The results are shown in Figure~\ref{cspace}, where the three quasars presented in this paper are shown along with random sources from the DPOSS survey and other BAL quasars drawn from the Junkkarinen, Hewitt, and Burbidge (1991) compilation, which are displayed using their colors as measured by DPOSS. The BAL quasars, including the peculiar BAL quasars presented herein, are differentiated based on their redshift: $1 \leq z < 2 $ ($\bigtriangleup$), $2 \leq z < 3$ ({\Large \sq}), and $z > 3$ ($\bigcirc$). The arrows on the three sources presented in this paper show the new location of these quasars after changing their redshift by $\pm 0.1$ (lower arrows result from shifting the spectrum to the red and the upper arrows result from shifting the spectrum to the blue). While not shown, we note that for comparison, unusual BAL quasars found within SDSS data (Hall \setal 2001) are considerably redder than normal quasars in $g-r$, and slightly redder in $r-i$.

While this figure demonstrates several things, the most important is the sensitivity of the location of these sources within color-space to their actual redshift. Two of these sources, PSS J0052+2405 and PSS J0141+3334, actually disappear into the stellar locus when their spectra are shifted blueward. This raises the possibility of additional peculiar sources lurking amongst the stars. In order to find them, data from multiple surveys (preferably covering different wavelength regions) will need to be federated. This results in a dataset that presents intense computational challenges due to the higher dimensional parameter space through which the algorithms must search. Through collaborations with computer scientists and statisticians, however, new algorithms and approaches (\eg Connolly \setal 2000) are being developed that provide possible solutions to this technical hurdle. 

An additional point of interest is the lack of differentiation between BAL quasars at different redshifts in this two-dimensional parameter space. This degeneracy could potentially be lifted with additional imaging data, and could prove to be an interesting application area for data mining algorithms. This is an area where medium band spectrophotometric surveys (\eg Combo-17, \url{http://www.mpia-hd.mpg.de/COMBO/combo\_index.html}) or multiwavelength data federation projects (\eg Brunner 2001) may provide fruitful input datasets.

In order to roughly estimate the rarity of these objects, we consider the
following.  These quasars have been found in the course of a search for
high-$z$ and type-2 QSOs, which covered the area of approximately $10^4$
$deg^2$.  At the brightness level of these sources, the spectroscopic
follow-up is nearly complete, in the corresponding region of the $gri$
color space.  Thus, the implied surface density of these objects, down
to this magnitude level and within the region of the color space searched,
is roughly $\sim 3 \times 10^{-4}$ $deg^{-2}$.
While we do not have any reliable way of estimating how many more such
objects exist outside our region of the color space, it is likely that 
the total (color-independent) surface density would be a factor of a
few higher.
While all 3 sources have $r < 19$ mag, our search was nearly complete down
to this flux level.  The cumulative surface density of all QSOs down to
such magnitude levels is approximately a few times $10$ $deg^{-2}$ (\eg Schneider \setal 2002).
Thus, we conservatively conclude that $\sim 10^{-4}$
of all QSOs belong to this peculiar subclass.

To conclude, in this paper we have presented three peculiar BAL quasars from the
Digitized Palomar Observatory Sky Survey: PSS J0052+2405, PSS J0141+3334, and PSS J1537+1227. 

PSS J0052+2405 is an FeLoBAL quasar at a redshift $z = 2.4512 \pm 0.0001$ with many species detected at the peak absorption redshift of $z = 2.406$,
including \SIiv, \SIii, \civ, \alii, \aliii, \NIii, \znii, \crii, \mnii,
and numerous \feii\ multiplets.

PSS J0141+3334 is a LoBAL quasar at $z = 3.005 \pm 0.005$ with several species seen in absorption, including \SIiv, \civ\ and \alii.  There is also narrow intervening absorption at lower redshifts from 
\feii\ and \mgii.

PSS J1537+1227 is a FeLoBAL quasar at  $z = 1.212 \pm 0.007$, and is an excellent target for high-resolution
spectroscopy, as it has very narrow absorption lines and may show absorption
from numerous useful diagnostic transitions.

The ability of large velocity width, high covering factor absorption to blanket
large regions of the spectrum in BAL quasars raises the possibility that 
quasar surveys based on blue colors have systematically missed such objects.
The true ranges of these and other BAL properties remains to be characterized,
most notably upper limits for the column densities of the various ions
which can be found in LoBAL outflows  (Hall \etal\ 2003). 

\acknowledgments

RJB acknowledges partial support from NASA ADP (NAG5-10885), NASA AISRP (NAG5-12000),  and the Fullam Award. PBH acknowledges financial support from Chilean grant FONDECYT/1010981 and a Fundaci\'{o}n Andes grant. The  authors gratefully acknowledge those of Hawaiian ancestry on  whose sacred mountain we are privileged to be guests.  Without their generous hospitality, none of the observations  presented would have been possible.  The processing of DPOSS and the production of the Palomar-Norris Sky Catalog (PNSC) on which this work was based was supported by generous grants from the Norris Foundation, and other private donors.  Some of the software development was supported by the NASA AISRP program.  We also thank the staff of Palomar Observatory for their expert assistance in the course of many observing runs.  Finally, we acknowledge the efforts of the POSS-II team, and the plate scanning team at STScI.




\clearpage

\begin{deluxetable}{rrrrrrrr} 
\tablecolumns{8} 
\tablewidth{0pc} 
\tablecaption{The coordinates (J2000), magnitudes, and redshifts for the three peculiar BAL quasars presented in this paper.} 
\tablehead{ 
\colhead{Source Name} & \colhead{RA}   & \colhead{DEC}    & \colhead{g} & 
\colhead{r}    & \colhead{i}   & \colhead{redshift}}
\startdata 
PSS J0052+2405\tablenotemark{a} & 00:52:06.8 & +24:05:39 & 19.37 & 18.08 & 18.87 & 2.45 \\ 
PSS J0141+3334\tablenotemark{b} & 01:41:32.9 & +33:34:24 & 20.88 & 18.70 & 18.54 & 3.01 \\
PSS J1537+1227\tablenotemark{a} & 15:37:41.8 & +12:27:44 & 18.67 & 18.26 & 18.52 & 1.21 \\
\enddata 
\tablenotetext{a}{The data for these quasars are publicly available from the DPOSS science archive: http://www.dposs.caltech.edu/}
\tablenotetext{b}{The magnitudes for this quasar are approximate, since this source does not lie within the well-calibrated area that has been released to the public.}
\label{dataTable}
\end{deluxetable}

\clearpage

\begin{figure}
%\epsscale{1.00}
\plotone{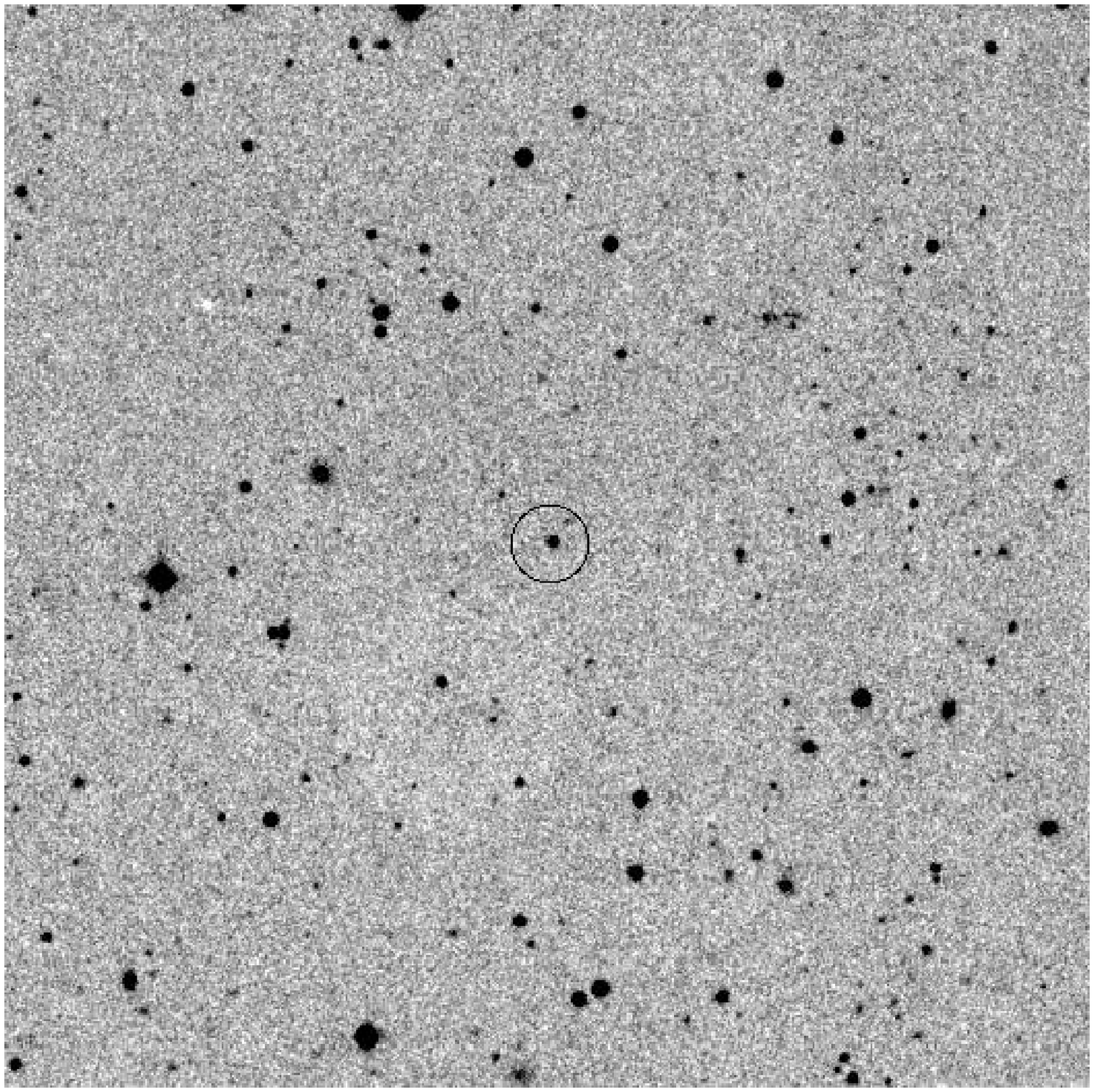}
\caption[]{A 512 x 512 pixel image ($\sim 8.5\arcmin$ x $8.5\arcmin$) in the $F$ band centered on PSS0052+2405, which is circled. In this image, North is up and East is right.}\label{pss0052}
\end{figure}

\begin{figure}
%\epsscale{1.00}
\plotone{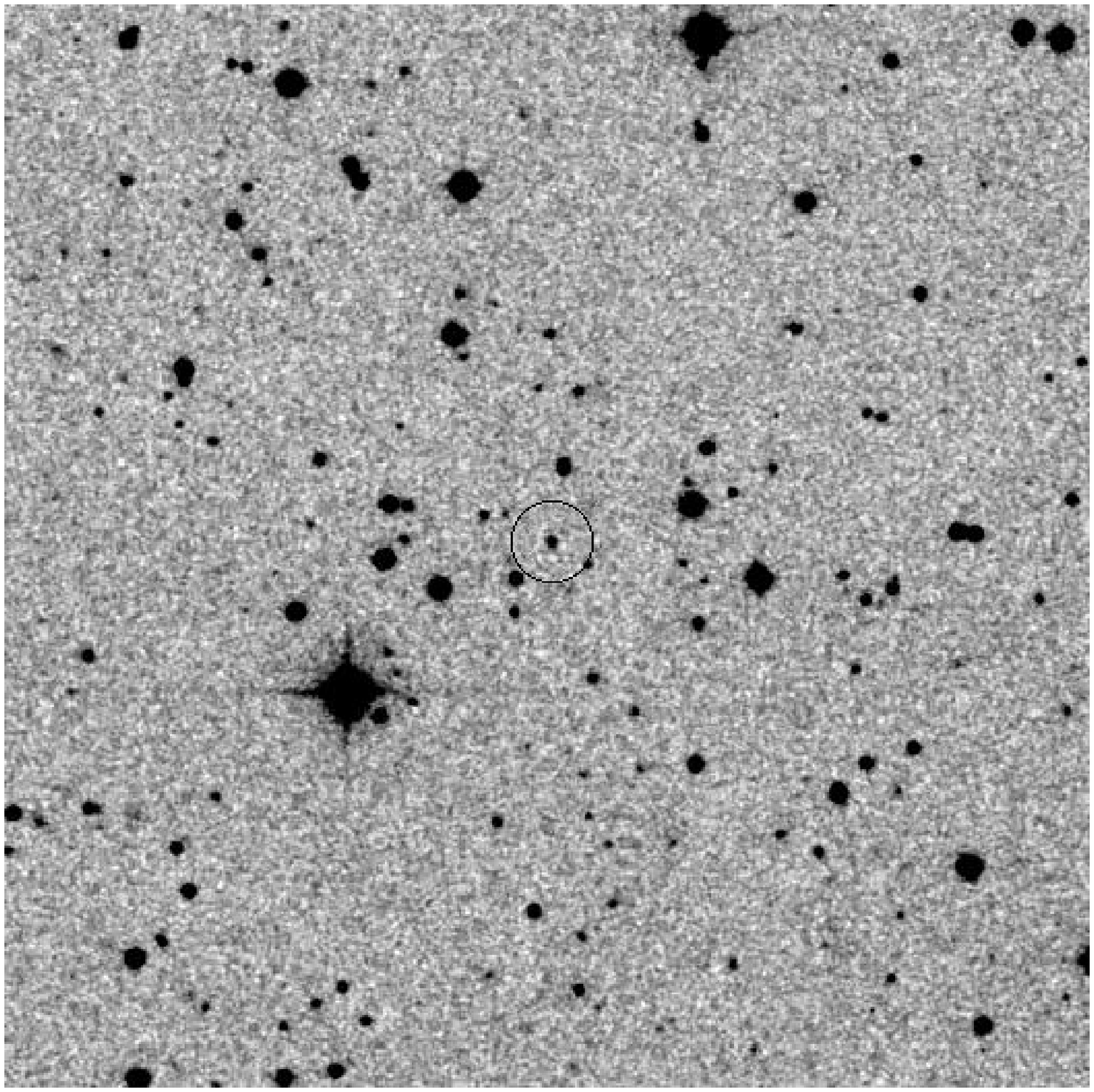}
\caption[]{A 512 x 512 pixel image ($\sim 8.5\arcmin$ x $8.5\arcmin$) in the $F$ band centered on PSS0141+3334, which is circled. In this image, North is up and East is right.}\label{pss0141}
\end{figure}

\begin{figure}
%\epsscale{1.00}
\plotone{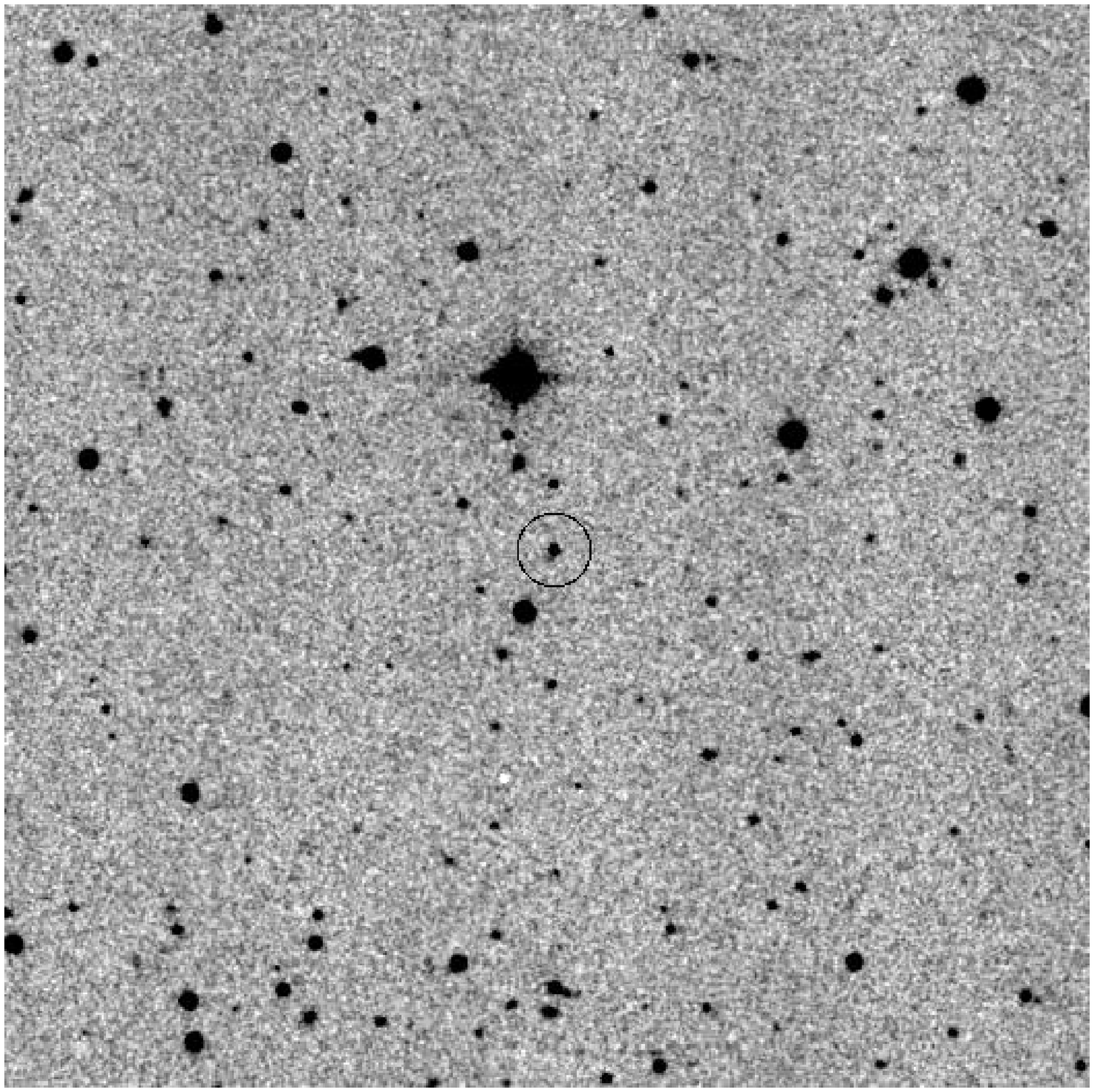}
\caption[]{A 512 x 512 pixel image ($\sim 8.5\arcmin$ x $8.5\arcmin$) in the $F$ band centered on PSS1537+1227, which is circled. In this image, North is up and East is right.}\label{pss1537}
\end{figure}

\begin{figure}
%\epsscale{1.00}
\plotone{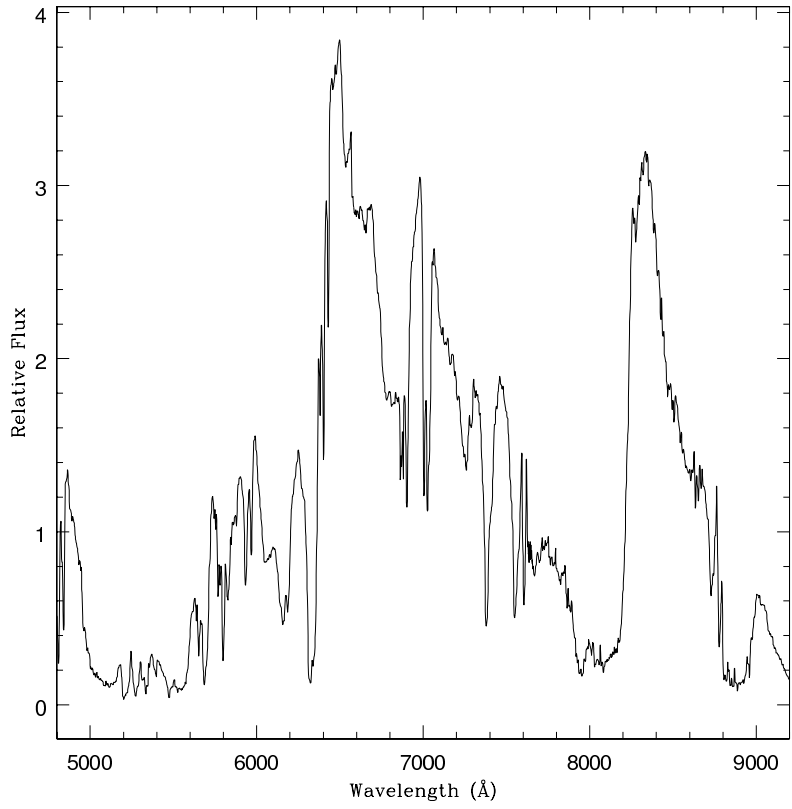}
\caption[]{The discovery spectrum ($f_{\nu}$ vs. $\lambda$) for PSS J0052+2405, obtained using ESI (Sheinis \setal 2002) on the Keck\,{\sc I} telescope.}\label{00522405Spec}
\end{figure}

\begin{figure}
%\epsscale{1.00}
\plotone{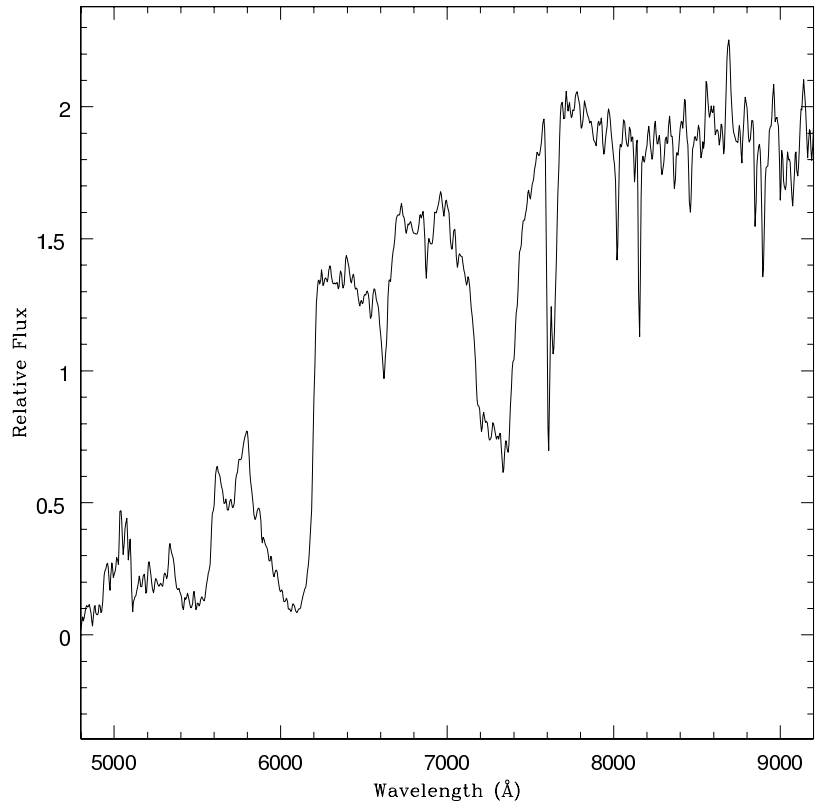}
\caption[]{The discovery spectrum ($f_{\nu}$ vs. $\lambda$)  for PSS J0141+3334, obtained using DBSP (Oke \& Gunn 1982) on the Hale 200-inch telescope.}\label{01413334Spec}
\end{figure}

\begin{figure}
%\epsscale{1.00}
\plotone{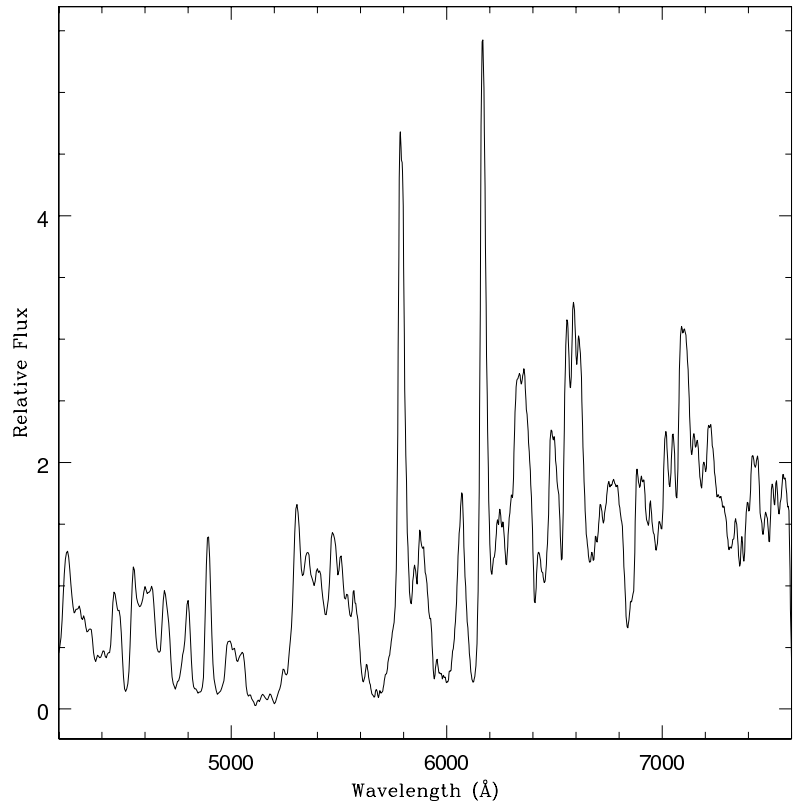}
\caption[]{The discovery spectrum ($f_{\nu}$ vs. $\lambda$)  for PSS J1537+1227, obtained using LRIS (Oke \setal 1995) on the Keck\,{\sc I} telescope.}\label{15371227Spec}
\end{figure}

\begin{figure}
%\epsscale{1.00}
\plotone{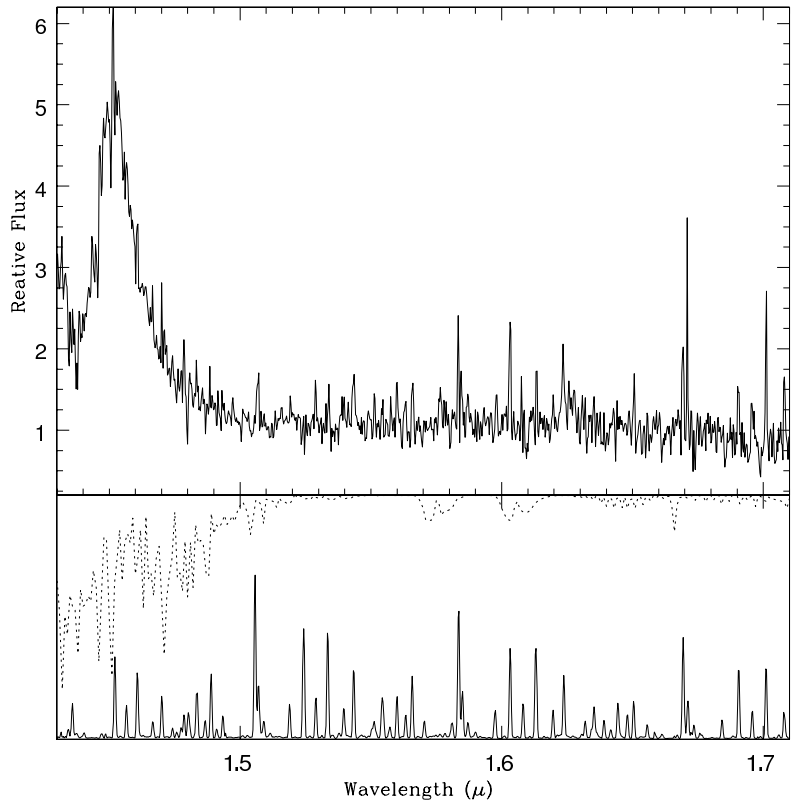}
\caption[]{The infra-red spectrum (relative flux vs. wavelength)  for PSS J1537+1227, obtained using NIRSPEC (McLean \etal\ 1998) on the Keck\,{\sc II} telescope. The top panel displays the actual spectrum, while the bottom panel displays the sky spectrum (solid line) and the relative atmospheric absorption (dotted line).}\label{pss1537_Hspec}
\end{figure}

\begin{figure}
%\epsscale{1.00}
\plotone{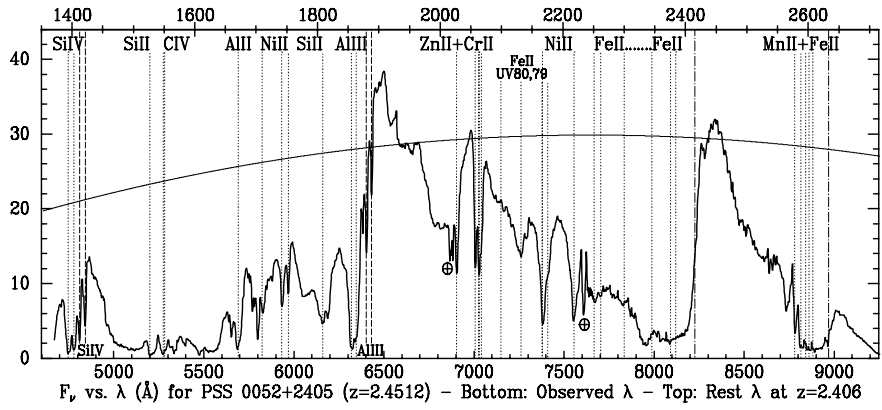}
\caption[]{
Spectrum of the FeLoBAL PSS J0052+2405, with observed wavelengths along the
bottom axis and rest-frame wavelengths at 
the peak absorption redshift of $z=2.406$ 
%our adopted systemic redshift of $z=2.4512$ 
along the top.  The vertical scale is $F_{\nu}$ in units of
$10^{-1}$ mJy.
Dotted lines indicate absorption at $z=2.4512$.
Dashed lines indicate absorption at $z=2.406$.
Dot-dash lines show the reddest excited \feii\ lines in the complexes
around 2400 and 2600\,\AA.
The thin solid line is the continuum fit adopted for calculation of the
balnicity index.
Telluric absorption is indicated by $\earth$ symbols.
}\label{fig0052opt}
\end{figure}

\begin{figure}
%\epsscale{1.00}
\plotone{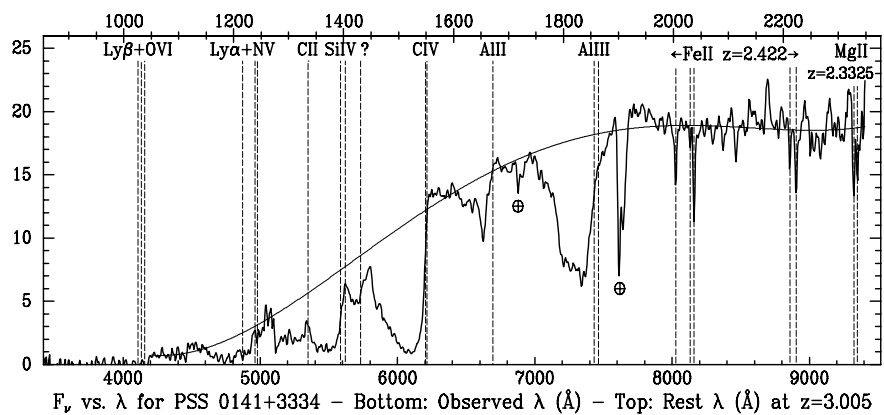}
\caption[]{
Spectrum of the LoBAL PSS J0141+3334, plotted in the same manner as
Figure~\ref{fig0052opt}.  Dashed lines are labelled with the transition and
redshift of the indicated absorption; transitions without a redshift label
are plotted at the systemic redshift of $z=3.005$.
}\label{fig0141opt}
\end{figure}

\begin{figure}
%\epsscale{1.0}
%\vspace{-3in}
\plotone{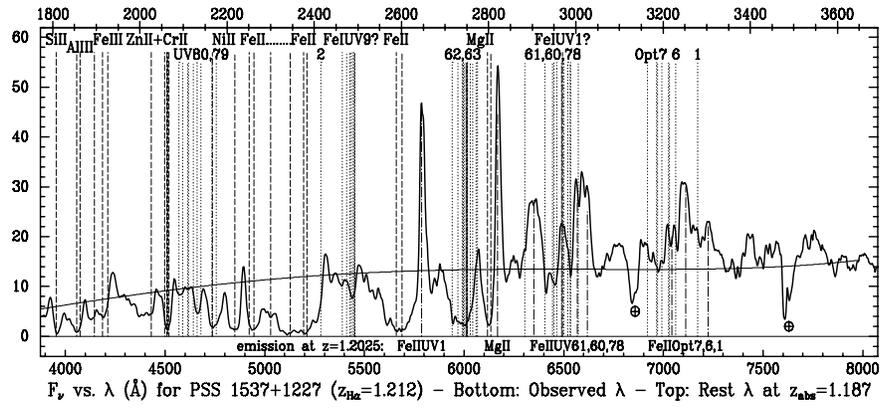}
\caption[]{
Spectrum of the FeLoBAL PSS J1537+1227, plotted in the same manner as
Figure~\ref{fig0052opt}.
Transitions in absorption are labeled across the top of the plot.
Dashed lines show confirmed transitions in absorption at $z=1.187$;
dotted lines show excited-state \feii\ transitions at that $z$.
%in absorption at $z=1.187$.
\feii\ ultraviolet (UV) and optical (Opt) multiplets are numbered
in the second line of labels across the top.
Transitions in emission are labeled across the bottom of the plot.
Dot-dashed lines show narrow emission at $z=1.2025$ from \mgii\
and various \feii\ multiplets.
%\feii\ and \mgii) and broad emission at $z=1.212$ (\fei+\feii).
}\label{fig1537opt}
\end{figure}

\begin{figure}
%\epsscale{1.00}
\plotone{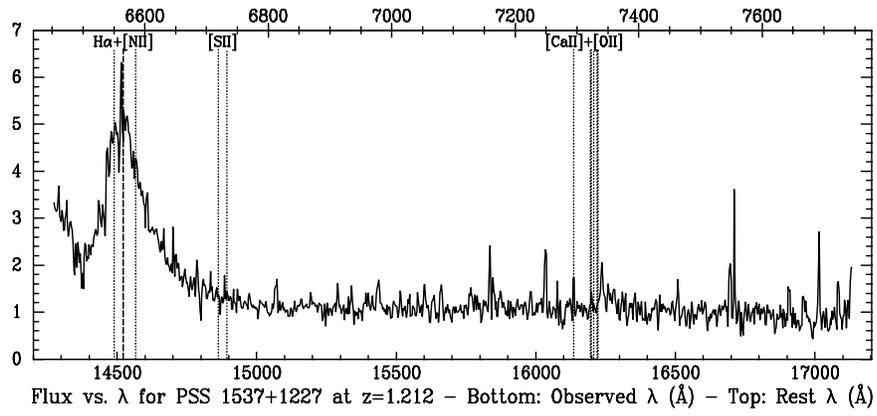}
\caption[]{
Infrared ($H$-band) spectrum of the FeLoBAL PSS J1537+1227.
Dashed lines show emission at the adopted systemic $z=1.212$.
The vertical scale is relative flux in arbitrary units.
}\label{fig1537ir}
\end{figure}

\begin{figure}
%\epsscale{0.75}
\plotone{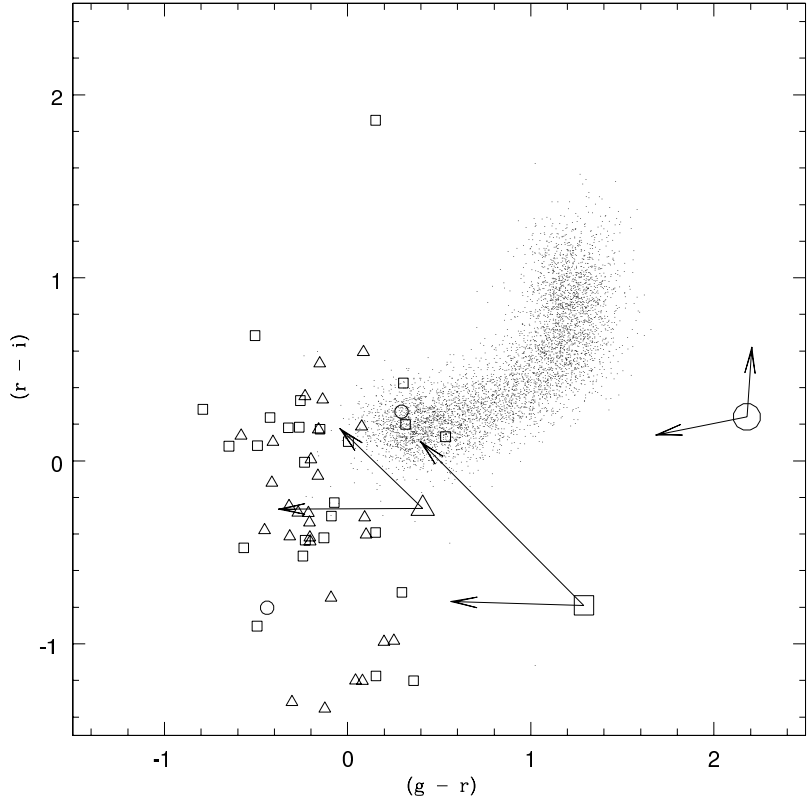}
\caption[]{The same color space and data as shown in Figure~\ref{target}, with the addition of other BAL quasars drawn from the Junkkarinen, Hewitt, and Burbidge (1991) compilation, which are displayed using their colors as measured by DPOSS. The three large symbols indicate the three quasars presented in this paper.  The BAL quasars, including the peculiar BAL quasars presented herein, are differentiated based on their redshift: $1 \leq z < 2 $ ($\bigtriangleup$), $2 \leq z < 3$ ({\Large \sq}), and $z > 3$ ($\bigcirc$). The arrows on the three sources presented in this paper show the new location of these quasars after changing their redshift by $\pm 0.1$ (lower arrows result from shifting the spectrum to the red and the upper arrows result from shifting the spectrum to the blue).

}\label{cspace}
\end{figure}

\end{document}